\documentclass[aps,pre,twocolumn,amsmath,amsthm,amssymb,showpacs]{revtex4}
\usepackage{graphicx}

\newtheorem{theorem}{Theorem}

\newtheorem{corollary}[theorem]{Corollary}

\begin{document}
 
\title{Understanding and Improving the Wang-Landau Algorithm}
\author{Chenggang Zhou}
\altaffiliation{{\em Current address:} Computer Science and Mathematics Division, Oak Ridge National Laboratory, P.O. Box 2008, MS6164, Oak Ridge TN, 37830-6164, U.S.A., Center for Simulational Physics, University of Georgia, Athens, Georgia 30602, U.S.A.}
\affiliation{ Department of Electrical Engineering, Princeton University,
Princeton, New Jersey 08544, USA}
\author{R. N. Bhatt} 
\affiliation{ Department of Electrical Engineering, Princeton University,
Princeton, New Jersey 08544, USA}
\date{\today}
 
\begin{abstract}
 
We present a mathematical analysis of the Wang-Landau algorithm, prove its convergence, identify
sources of errors and strategies for optimization. In particular, we found the histogram increases uniformly with small fluctuation after a stage of initial accumulation, and the statistical error is found to scale as $\sqrt{\ln f }$ with the modification factor $f$.
This has implications for strategies for obtaining fast convergence.
\end{abstract}

\maketitle
 
The Wang-Landau(WL) algorithm \cite{WangLandau} has been applied to a number of interesting
problems~\cite{WangLandau,Okabe, Shell, Pablo, Troyer, Tomita}.
It overcomes some difficulties in other Monte Carlo algorithms such as critical slowing 
down, and long relaxation times due to frustration and complex energy terrain. 
Similar to the Metropolis algorithm, it is a generic algorithm, independent on the details 
of the physical system. Many methods have been suggested to improve the algorithm for certain
types of systems~\cite{Kawashima, Schulz02, Schulz03}.
The same mechanism also appears in the recent research of molecular dynamics simulations~\cite{Wu04}. 
Among the studies to characterize and improve the efficiency of the algorithm, 
Dayal {\em et al.}\cite{Dayal} shows the WL algorithm considerably reduces the tunneling time,
and Trebst {\em et al.}\cite{Trebst04} proposed an algorithm that performs better in terms of
tunneling time.
However, the WL algorithm has been used as an empirical method. Many important questions 
still remain unanswered: 
(i) How is flatness of the histogram related to the accuracy; 
(ii) what is the relation between the modification factor and error; 
and (iii) how does the simulation actually find out the density of states?
The convergence of the WL algorithm should be guaranteed by a generic principle,
in the same sense as the detailed balance assures the convergence of the Metropolis algorithm. However, the
WL algorithm is different from the Metropolis algorithm, since it is not a Markov process. 
 
In this paper we present our study of this algorithm from an analytical approach, 
and try to answer those questions raised above. 
Our analysis provides a proof of the convergence of the method, estimation for
the errors and the computational time, along with some strategies for optimization 
and parallelization. 
 
The goal of the WL algorithm is to accumulate knowledge about $\rho(E)$ during a 
Metropolis-type MC sampling. 
The Metropolis-type random walk is characterized by an acceptance ratio 
$\min\{1, g(E_j)/g(E_i)\}$, 
where $g(E)$ is a function of energy, similar to the Boltzman factor in the usual 
Metropolis algorithm. $E_i$ and $E_j$ refer to energies before and after this transition.
the acceptance ratio biases the free random walk and produces a final histogram $h(E)$,
which is related to the equilibrium distribution of the unbiased random walk $\rho(E)$ by
$\rho(E)g(E)=h(E)$, provided that both sides of the identity are normalized.
This identity is essentially a result of detailed balance.
The WL algorithm divides $g(E_j)$ by a modification factor $f$ after each transition, 
expecting $g(E)$ to converge to $1/\rho(E)$ and the histogram $h(E)$ to be flat.
 
$\rho(E)$ is {\it a priori} unknown in the simulation. We begin our analysis 
by clarifying the relevant parameters. Suppose the phase space of our physical model is
divided into $N$ macroscopic states with density (number of microscopic configurations) 
$\rho_i>0$ for macroscopic state $i$ ($i=1,2\cdots N$). 
(These macroscopic states could be labeled by energy, magnetization, 
or other macroscopic variables). Each microscopic configuration in the phase space 
uniquely belongs to one macroscopic state. 
The histogram $h_i(t)$ with $1\leq i\leq N$ is defined as the number 
of visits of each macroscopic state before $t$\textsuperscript{th} step of the simulation. 
Initially $h_i(1)=0$, after macroscopic state $k$ is visited at time $t$, 
$h_i(t+1) = h_i(t)+\delta_{ik}$. 
In the original implementation~\cite{WangLandau}, one record is inserted into the histogram every Metropolis trial flip.  However, in addition to the modification factor, 
we have a second tunable parameter of the algorithm, separation $S$ between successive 
records in the histogram, defined as the number of trial flips (random steps) that precede 
each increment of the histogram. $S$ steps of random walk should be regarded as a single 
transition from initial macroscopic state $i$ to the final macroscopic state $j$. 
Namely, the transition from $i$ to $j$ includes $S$ trial flips, each trial flip makes a 
transition from macroscopic state $k_{n-1}$ to macroscopic state $k_n$ ($k_0 = i$ and $k_S=j$) 
with acceptance ratio $\min\{1, \exp[\ln f(h_{k_{n-1}}(t) - h_{k_{n}}(t))] \}$. 
At time $t$, macroscopic state $k$ has a probability $p_k(t)$ to be picked out by the simulation. 
$p_k(t)$ is normalized so that vector $p(t) \in V_N$, where $V_N = \{x \in [0,1]^N, \sum_{k=1}^N x_k = 1 \}$ is an N-dimensional simplex. 
In the following derivation, we assume $S$ is larger than the autocorrelation length of the random walk, then $i$ and $j$ can be considered as independent random macroscopic states. The effect of autocorrelation will be discussed later. Although this assumption differs from real simulations, it is reasonable since the particular model under study is not specified. It is also asymptotically accurate for any model when $S$ is large enough. With this assumption, the probability distribution $p(t)$ 
has an explicit expression,
\begin{equation}
\label{prob}
  p_i(t) = Z(t)^{-1} (\rho_i/\theta_i)f^{-h_i(t)},
\end{equation}
where $Z(t) = \sum_{k=1}^N(\rho_k/\theta_k)f^{-h_k(t)}$ is the normalization constant, 
and without loss of generality, we insert an initial guess $\theta_i$ into the simulation. 
In fact $f^{h_i(t)}$ serves as a guess for the simulation after $t$. Similarly the 
simulation can start from some ``existing knowledge'' represented by $\theta_i$. If nothing is 
known about the density of states, then we start with $\theta_i=1$. With the probability 
distribution Eq.~(\ref{prob}), the macroscopic state $k$ has the probability $p_k(t)$ to be 
picked out in the next step. Once this happens, $p(t)$ is changed to $p(t+1)$ by
\begin{equation}
  p_i(t+1) = p_i(t)f^{-\delta_{ik}}/[1-p_k(t)+p_k(t)f^{-1}].
\end{equation}
Here the Kronecker-$\delta$ only suppresses $p_k(t)$ by a factor of $f$ and the denominator 
comes from the change in $Z(t+1)$ such that the normalization of probability is preserved. We
point out that the evolution of $p(t)$ is a Markov process, although the WL algorithm 
is not, because it makes references to its entire history.

We will prove that the $p(t)$ is attracted to the vicinity of uniform distribution ($p^{(0)}_i = 1/N$) in the simulation. For this purpose, we define a measure of the difference between $p(t)$ and the uniform distribution $p^{(0)}$ by
\begin{equation}
\mu(t) = N\ln N + \sum_{i=1}^N \ln p_i(t).
\end{equation}
One can check that $\mu(t)\leq 0$ and $\mu(t) = 0$ only when $p(t) = p^{(0)}$. After the macroscopic state $k$ is picked out, $\Delta \mu(t) = \mu(t+1)-\mu(t) $ is given by
\begin{equation}
\label{dmu}
 \Delta \mu(t)= -\ln f - N \ln [1-p_k(t)+p_k(t)f^{-1}].
\end{equation}
Obviously $\Delta \mu(t)>-\ln f$, and when $p_k(t)>{1-f^{-1/N} \over 1-f^{-1}} \approx \ln f / N (1-f^{-1})$ (approximately $1/N$ when $f \rightarrow 1$), $\Delta \mu(t)$ is always positive. This shows that $\mu(t)$ increases, when the simulation picks out macroscopic states with probabilities above average. However there is a probability for $\mu(t)$ to decrease, in particular, at the center of attraction ($p(t) = p^{(0)}$), $\Delta \mu(t)$ is negative. As a result, rather than converging to the uniform distribution, $p(t)$ is expected to either fluctuate around uniform distribution, or go away from it. Actually the second situation does not happen. 
To prove this, we first show that the expectation value $\mathbf{E}_{p(t)} \{ \Delta \mu(t) \}$ (averaged over all possible moves) has a lower bound determined by $p(t)$:
Using Eq.~(\ref{dmu}), the expectation value becomes
\[
 \mathbf{E}_{p(t)} \{ \Delta \mu(t) \} = -\ln f
+ N \sum_{k=1}^N p_k(t)\ln {1 \over 1-p_k(t)(1-f^{-1})}.
\]
Since $0<p_k(t)(1-f^{-1})<1$, we use the inequality $\ln (1-x)^{-1} > x$ 
for $x \in (0,1)$ to give a lower bound for the logarithm, which turns out to be
\[ 
\mathbf{E}_{p(t)} \{ \Delta \mu(t) \} > -\ln f+ N (1-f^{-1})\sum_{k=0}^N p_k^2(t).\]
Typically $p_k(t)$ is of order $1/N$, where $N$ is a large integer, so this lower bound is very close to the actual value. 
This lower bound can be further expressed in terms of the Euclidean distance between $p(t)$ and $p^{(0)}$, since  $\|p(t)-p^{(0)}\|^2 = \|p(t)\|^2 -1/N$, due to the normalization of $p(t)$.
Therefore we have:

\begin{theorem}
\label{t1}
If $ \|p(t)-p^{(0)}\|^2 = N^{-1}[(1-f^{-1})^{-1}\ln f-1]+\epsilon$, with $\|\cdots\|$ being the Euclidean distance, the expectation value of $\Delta \mu(t)$ averaged over $N$ possible moves is bounded from below
$ \mathbf{E}_{p(t)} \{ \Delta \mu(t) \} > N (1-f^{-1} )\epsilon$.
\end{theorem}

Theorem~\ref{t1} states that for a probability distribution $p(t)$ outside the N-dimensional 
sphere $B_\epsilon$ defined by its condition, $\mu(t)$ always has a tendency to increase. 
Next we consider an ensemble of simulations, whose $p(t)$ has a certain distribution at time $t$, $F_t(p)$. The ensemble averaged $\mu(t)$ is defined as $\left<\mu(t)\right> = \int F_t(p) \mu(p) dp $. (note in the integrand we treat $\mu$ as a function of $p$ instead of time $t$.) We want to show that the evolution of $F_t(p)$ brings every simulation in the ensemble into the sphere $B_{\epsilon}$. Define $D(p,p')$ as the probability of bringing distribution $p'$ to $p$ after one step. Obviously, $F_{t+1}(p) = \int_{V_N} D(p,p')F_t(p')dp'$, where the integral over $p'$ is restricted to the simplex $V_N$. 

We can express the ensemble average of $\mu(t+1)$, $\left<\mu(t+1)\right> = \int_{V_N} F_{t+1}(p)\mu(p)dp$, with $F_t(p)$:
$\left<\mu(t+1)\right> = \iint_{V_N} \mu(p)D(p,p')F_t(p')dp'dp$. As a result of Theorem \ref{t1}, if we assume at time $t$, every simulation is outside $B_\epsilon$, i.e. $p' \not \in B_\epsilon$, then $\int_{V_N} \mu(p)D(p,p')dp > \mu(p')+ N(1-f^{-1})\epsilon$. Therefore $\left<\mu(t+1)\right> > \int_{V_N} [\mu(p')+ N(1-f^{-1})\epsilon ] F_t(p')dp' = \left<\mu(t)\right> + N(1-f^{-1})\epsilon$, which is the following corollary:

\begin{corollary}
\label{c2}
If the distribution $F_t$ does not enter $B_\epsilon$, i.e. ${\rm supp}(F_t) \cap B_\epsilon = \emptyset $, then $\left<\mu(t+1)\right> > \left<\mu(t)\right> + N(1-f^{-1})\epsilon$.
\end{corollary}

This result implies that $\left<\mu(t)\right>$ increases at least linearly as the simulation goes.
However, if ${\rm supp}(F_t)$ is always outside $B_\epsilon$, $\mu(t)$ is therefore bounded from 
above by the maximum value of $\mu(p)$ on the boundary of $B_\epsilon$. This contradiction tells us that 
part of the ensemble has to be pushed into $B_\epsilon$ so that 
${\rm supp}(F_t) \cap B_\epsilon \neq \emptyset $. We can exclude those parts of the ensemble 
already inside $B_\epsilon$, and apply the same inference to the remaining part of the ensemble 
which is still outside $B_\epsilon$. 
The conclusion is, no matter where the simulation starts, $p(t)$ sooner or later goes into $B_\epsilon$. Once $p(t)$ is in the vicinity of $p^{(0)}$, it is unlikely to escape because $\Delta \mu(t)$ has a lower bound $\Delta \mu(t)>-\ln f$. If after a certain step $p(t)$ moves outside $B_\epsilon$, we can immediately use Corollary \ref{c2} to show that it is attracted
back into $B_\epsilon$.
It is this attraction towards $p^{(0)}$ that reduces the tunneling time of the WL algorithm \cite{Dayal}. When $N(1-f^{-1})\ll 1$, this attraction is weak, which explains why Ref.~\cite{Dayal} finds the tunneling time of the WL algorithm saturates when $f$ is less than a critical value determined by the system size.

\begin{figure}[b]
 \includegraphics[width=0.75\columnwidth]{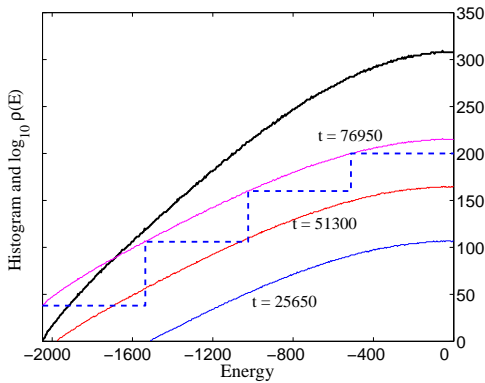}
 \caption{Snapshots of the histogram of a single random walker on a two-dimensional 32 by 32 Ising model.
Three thin curves are histograms of three sequences of lengths labeled in the figure. The thick dark line
is the $\rho(E)$ calculated from the last histogram ($t=76950$), which overlaps with the exact $\rho(E)$ within the accuracy of the width of the line on this figure. The length of the sequence is just the area under the histogram. The dashed staircase indicates a possible guess for $\rho(E)$ with four energy intervals (see text).}
\label{fig2}
\end{figure}

When $p(t)$ is trapped near $p^{(0)}$, the histogram shows a uniform growth with fluctuation. 
$h_i(t) =  \log_f (\rho_i / \theta_i) + { t / N}+ r_i(t)$, where $r_i(t)$ is a random number with zero mean. The approximate density of states is measured from the histogram by $\rho'_i(t) = K \theta_if^{h_i(t)}$, where $K$ is a proper normalization constant.
Figure \ref{fig2} shows three snapshots of the
histograms calculating the density of states of a two-dimensional Ising model on a lattice of size 32 by 32 with periodic boundary conditions. We have used $f = e^4$ in this simulation to reveal the fluctuation of the histogram in Fig.~\ref{fig2}. The simulation visits energies (macroscopic states) with high density of states first, then extends the histogram to the whole spectrum. Once the whole spectrum is visited, the histogram grows uniformly with small fluctuation.
Two important observations follow the results above:
(i) A flat histogram is not required for convergence; the histogram is ready for calculation of $\rho'$ when it reaches a threshold $h_i(t) \gg \sigma_{r_i}$, for all $i$'s, where $\sigma_{r_i}$ denotes the standard deviation of $r_i(t)$. (ii) The statistical error  can be reduced by averaging over multiple results obtained with the same $f$, as well as reducing $f$. A proper estimation of the statistical error is from the condition of Theorem \ref{t1}. Since $p(t)$ fluctuates around a ball centered at $p^{(0)}$ 
of radius $ N^{-1/2}\pi(f) = \sqrt{N^{-1}[(1-f^{-1})^{-1}\ln f -1]}$, we use $\pi(f)$ to give an reasonable estimation of the statistical error. $p(t)$ is related to $\rho'(t)$ by $p_i(t) = C \rho_i \rho'^{-1}_i(t)$, where $C$ is a constant for all $i$'s. Plugging this into $\|p(t)-p^{(0)}\| \approx N^{-1/2}\pi(f)$, we arrive at
\begin{equation}
\sqrt{ {1 \over N} \sum_{i=1}^N \left( {NC \rho_i \over \rho'_i(t)} -1\right)^2}
\approx \pi(f),
\end{equation} 
where $NC$ is a constant allowed by the WL algorithm, which can be absorbed into $\rho'_i(t)$. The left side of this equation is the standard deviation of $\rho_i / \rho'_i(t)$. So an appropriate estimation of the typical relative error $\delta \rho'_i /\rho'_i $ is $\pi(f)$, which scales as $\sqrt{\ln f }$ when $f$ is close to 1.
(The difference between the expression here $(\rho'-\rho)/\rho'$ and the standard definition of relative error $(\rho'-\rho)/\rho$ is negligible, when the error is small.) Thus, we expect the fluctuation in the histogram to be
$\sigma_{r_i} \approx \pi(f)/\ln f \sim 1/\sqrt{\ln f}$, because $d\rho'_i(t)/dh_i(t) = \rho'_i(t)\ln f$. This has been recently confirmed by numerical tests~\cite{Lee04}.
Our strategy for a single iteration simulation is to run
until a minimum number of visits (at least $1/\sqrt{\ln f }$) have been accumulated for each 
macroscopic state,
followed by measurements separated by a short simulation which decorrelates $r_i(t)$. Usually
$1/\sqrt{\ln f}$ visits on each macroscopic state is enough. 
With $K$ measurements, the statistical error in $\ln \rho'_i(t)$ is reduced to 
$\sqrt{\ln f / K}$. The total number of records in the histograms is thus at least:
\begin{equation}
\label{step}
\sum_{i=1}^N h_i(t) \approx \sum_i \log_{f}{\rho_i \over \min_i\{\rho_i\}}+ 
{ NK \over \sqrt{\ln f}}.
\end{equation}
The first term represents the number of records for the simulation to reach every macroscopic state. This term
occupies the bulk of the histograms in Fig.~\ref{fig2}. The second term of Eq.~(\ref{step}) represents the cost of $K$ uncorrelated measurements. The measurements can be parallelized on a number of processors. The dashed staircase in Fig.~\ref{fig2} shows what happens if the energy is divided into 4 equal intervals, as Ref.~\cite{WangLandau} did for parallelization. As the spectrum is divided into four intervals assigned to four separate simulations, an interval with high density of states does not have to wait for those with low density of states to be visited. The histogram represented by the first term of Eq.~(\ref{step}) is reduced to the area of 4 triangles bounded by the staircase and the last histogram above it in Fig.~\ref{fig2}. In terms of saving total 
computational time, an equivalent strategy is to use the staircase as an initial guess 
$\theta_i$. Thus, four triangles are filled simultaneously, equivalent to doing 4 
simulations sequentially. Dividing the energy range causes boundary errors~\cite{Schulz03},
while a good initial guess of the functional form of the histogram does not have this problem.

Assuming $N$ is roughly proportional to the total number of degrees of freedom, to the logarithm of maximum density of states, and to the number of MC steps to generate an uncorrelated visit, the cost of CPU time for the first term in Eq.~(\ref{step}) is of order $O(N^3)$, while the cost of the second term is of order $O(N^2)$. (Logarithmic corrections might be present.) If we use a proper guess $\theta_i$ to begin the simulation, the CPU time cost for the first term can be substantially reduced to $O(N^2)$.

\begin{figure}[htbp] 
\includegraphics[width=0.75\columnwidth]{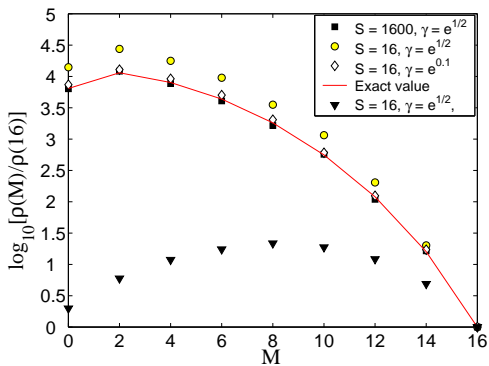}
\caption{$\rho(M)$, density of states for  magnetization $M$ of 16 Ising spins 
normalized to $\rho(M=16)$. 
The solid line connects the exact values, symbols were obtained with different 
parameters $f$ and $S$ as explained in the text. Data was averaged over 100 measurements, 
so the statistical errors are smaller than the symbols. }
\label{fig3}
\end{figure}

Now we discuss the effect of insufficient $S$ which introduces autocorrelation between 
successive records in the histogram. At first sight, the total number of steps given by 
Eq.~(\ref{step}) is considerably reduced by using a large $f$.
Multiple measurements also reduces the statistical error more quickly than reducing the value 
of $f$. So a small $f$ seems to be unnecessary. 
However, there {\bf are} {\it systematic} errors due to the correlation between adjacent 
records in the histogram when $f$ is not small, or the separation $S$ not large enough. 
We illustrate this systematic error in Fig.~\ref{fig3}, which shows the total density of states 
for a fixed {\it magnetization} $M$, $\rho(M)$, of the Ising model on a $4\times 4$ lattice, 
for which the exact result is: 
$\log_{10}\rho(M) = \log_{10}C_{16}^{8+M/2} + (1-\delta_{0M})\log_{10}2$. 
States with $M$ and $-M$ are grouped in the same macroscopic state for a better statistics, 
so $M$ is restricted to be a non-negative even integer. 
We demonstrate the effect of this correlation in Fig.~\ref{fig3} by showing the result of an 
extremely biased scheme that restores all 16 spins to total alignment after each record. 
As expected, the result (shown as downward triangles) is biased towards $M=16$. The 
simulation actually calculates the probability of reaching state $M$ starting from $M=16$ 
within 16 trial flips, which deviates from the correct density of states. 
On the other hand, The accuracy of data shown as squares indicates 
that the desired probability distribution Eq.~(\ref{prob}) is produced after 1600 trial flips. 
The data plotted with diamonds for a smaller $f$ show that the systematic error is also 
reduced by letting $f\rightarrow 1$, or equivalently, the minimum $S$ required to 
eliminate the autocorrelation decreases with $f$. As an extreme case, if each trial 
flip is recorded ($S=1$), for the Ising model of Fig.~\ref{fig2}, the histogram always grows 
quickly near $E=0$ and propagates to higher or lower energies very slowly. 
The systematic error due to the correlation is revealed, when the statistical error is reduced 
by multiple measurements with a single $f$.  At this point,
either a smaller $f$, or larger $S$ is necessary to improve the accuracy. 
 
To summarize, we have given an proof of the convergence of the WL algorithm, and analyzed the
sources of errors and optimization strategies. 
We find: (i) The density of states is encoded in the average histogram; 
(ii) The fluctuation of the histogram, proportional to $1/\sqrt{\ln f}$, where $f$ 
is the modification factor, causes statistical error, which can be reduced by averaging 
over multiple $\rho'(t)$. (iii) The correlation between adjacent records in the histogram 
introduces a systematic error, which is reduced by small $f$, and also by minimizing the 
correlation, e.g. using large $S$, or cluster algorithms \cite{Swedsen}.
These findings suggest that numerical simulations can start with a large $f$, e.g. $e^4$, 
and then reduce $f$ in large steps in each stage, e.g. divide $\ln f$ by a factor of 10. 
Multiple measurements can be made in the final stage to reduce the statistical error effectively. 
However, results calculated with a single pair of $f$ and $S$ is prone to systematic error. 
One can extrapolate results calculated with different $f$ and $S$ to $f=1$ or $S=\infty$
to eliminate this error. If the error is believed to be small enough, 
one can also reduce $f$ to 1 directly, which results in a 
histogram proportional to $\rho_i / \rho'_i(t)$, which is the ratio of the true density of states to the numerical results.

This research was supported by NSF DMR-0213706. We acknowledge discussions with D. P. Landau, D. A. Huse, M. S. Shell, and D. Guo.

\end{document}